\documentclass[prd,nofootinbib,nobibnotes,twocolumn]{revtex4}
\usepackage{epsf,epsfig}
\usepackage{amssymb,amsmath,amsfonts}

\newcommand{\CL}{\mathcal{L}}
\newcommand{\CO}{\mathcal{O}}
\newcommand{\CN}{\mathcal{N}}

\newcommand{\R}{\mathbb{R}}

\newcommand{\gym}{g_{\rm YM}}
\newcommand{\ads}{\mbox{AdS}}

\newcommand{\nn}{\nonumber}

\newcommand{\spa}{\ , \ \ }

\newcommand{\tr}{\mathop{{\rm Tr}}}

{}{}
{}{}
\def\vereq#1#2{\lower3pt\vbox{\baselineskip1.5pt \lineskip1.5pt
\ialign{$\m@th#1\hfill##\hfil$\crcr#2\crcr\sim\crcr}}}
\makeatother

\def\be{\begin{equation}}
\def\ee{\end{equation}}
\def\bea{\begin{eqnarray}}
\def\eea{\end{eqnarray}}

\allowdisplaybreaks

\usepackage[breaklinks=true]{hyperref}
\hypersetup{   colorlinks=true,  citecolor=blue, linkcolor=red, urlcolor=blue}

\begin{document}


\title{Interacting Giant Gravitons from Spin Matrix Theory} 
        
\author{Troels Harmark} 
\address{\vspace{0.2cm} 
The Niels Bohr Institute, Copenhagen University, Blegdamsvej 17, DK-2100 Copenhagen \O, Denmark \vspace{0.2cm} \\
{\small \tt harmark@nbi.ku.dk}  \\
}
\begin{abstract}
Using the non-abelian DBI action we find an effective matrix model that describes the dynamics of weakly interacting giant gravitons wrapped on three-spheres in the AdS part of $\ads_5\times S^5$ at high energies with two angular momenta on the $S^5$. In parallel we consider the limit of $\CN=4$ super Yang-Mills theory near a certain unitarity bound where it reduces to the quantum mechanical theory called $SU(2)$ Spin Matrix Theory. We show that the exact same matrix model that describes the giant gravitons on the string theory side also provides the effective description in the strong coupling and large energy limit of the Spin Matrix Theory. 
Thus, we are able to match non-supersymmetric dynamics of D-branes on $\ads_5\times S^5$ to a finite-$N$ regime in $\CN=4$ super Yang-Mills theory near a unitarity bound.

\end{abstract}
\maketitle



\section{Introduction and summary}

The AdS/CFT correspondence asserts that $SU(N)$ $\CN=4$ super Yang-Mills theory (SYM) is dual to type IIB string theory on $\ads_5 \times S^5$ with $N$ units of Ramond-Ramond five-form flux \cite{Maldacena:1997re}. 
As consequence of this holographic duality one should be able to see how space, time and gravity emerge from a quantum theory. However, this requires that one is able to connect the two sides of the correspondence quantitatively. 

In the planar regime $N=\infty$, where planar $\CN=4$ SYM is dual to tree-level type IIB string theory on $\ads_5\times S^5$, such a quantitative connection has been found in the form of an integrable spin chain \cite{Beisert:2010jr}. This provides a unifying framework for the two sides of the correspondence, enabling one to interpolate the full spectrum of the two theories from weak 't Hooft coupling $\lambda = \gym^2 N \ll 1$, where $\CN=4$ SYM is a good description,  to strong coupling $\lambda \gg 1$, where type IIB string theory is a good description, and vice versa.

Can one find a unifying framework beyond the planar regime that generalizes the spin chain theory? On the face of it, this seems highly difficult since presumably one does not have the integrability symmetry for finite $N$. The proposal of \cite{Harmark:2014mpa} is to take a different limit that accesses a regime that includes finite $N$ effects. The starting point is to consider one of the unitarity bounds $E \geq J$ of $\CN=4$ SYM where $E$ is the energy and $J$ is a linear combitation of charges of $\CN=4$ SYM. The limit then takes $E - J\rightarrow 0$ keeping $(E-J)/\lambda$ finite. In such a limit, $\CN=4$ SYM simplifies greatly and is effectively described by a quantum mechanical theory called Spin Matrix Theory (SMT) \cite{Harmark:2014mpa}. One can equivalently take the limit in the grand canonical ensemble by approaching a zero-temperature critical point. 

In \cite{Harmark:2014mpa} $SU(2)$ SMT is studied as the simplest non-trivial example. This is associated to the bound with $J=J_1+J_2$ where $J_1$ and $J_2$ are two R-charges of $\CN=4$ SYM. We found two tractable regimes in which one can interpolate from weak to strong coupling. One is analogous to the planar regime, and is described by the ferromagnetic $XXX_{1/2}$ Heisenberg spin chain. In the other regime $N$ is finite, but one considers $J$ so high that $SU(2)$ SMT is described by a classical matrix model. 

In this paper we find the first direct evidence that one is able to use Spin Matrix Theory to interpolate non-supersymmetric finite-$N$ effects from $\CN=4$ SYM to type IIB string theory on $\ads_5\times S^5$. Namely, by studying the non-abelian DBI action for $k$ giant gravitons wrapped on three-spheres in $\ads_5$, we find the exact same classical matrix model on the string theory side, focussing on the leading contribution to the giant gravitons. The matrix model we find is
\begin{eqnarray}
\label{string_MM}
E  &=&  \frac{1}{2} \tr ( P_1^2 + P_2^2 + X_1^2 + X_2^2) \nn \\ &&
 - \frac{g_s}{8\pi}\tr \Big(  [ X_1,X_2]^2  + [P_1,P_2]^2   + [X_1,P_1]^2 
 \nn \\ &&
 + [X_2,P_2]^2 + [X_1,P_2]^2 + [X_2,P_1]^2   \Big)
\end{eqnarray}
subject to the Gauss constraint $[X_1,P_1]+[X_2,P_2]=0$. This is valid for high energies $J \gg kN$ and to first order in the string coupling $g_s$. $X_1,X_2,P_1,P_2$ are $k$ by $k$ Hermitian matrices. Using the AdS/CFT dictionary $\gym^2 = 4\pi g_s$ this is seen to be same matrix model as found from $SU(2)$ SMT \cite{Harmark:2014mpa}. $J$ corresponds to the first line in Eq.~\eqref{string_MM}. 

The above described result means that we are able to match non-supersymmetric dynamics of D-branes on $\ads_5\times S^5$ to a finite-$N$ regime in $\CN=4$ SYM near a unitarity bound.
Previous results on giant gravitons have considered supersymmetric configurations \cite{Kinney:2005ej,Biswas:2006tj,Mandal:2006tk} or have focussed on matching the dispersion relation for open strings stretched between giant gravitons \cite{Balasubramanian:2002sa}.

The matrix model \eqref{string_MM} is derived for $1/N \ll g_s \ll 1$ while the corresponding result from $SU(2)$ SMT assumes $g_s \ll 1/N$. One could wonder why one should expect an exact match for the matrix models found in these two different regimes, $i.e.$ whether there is a order-of-limits issue. An analogous question can be posed in the planar regime of $SU(2)$ SMT where one finds the same Landau-Lifshitz model action from the $\CN=4$ SYM and type IIB string theory sides to first order in $g_s$ \cite{Kruczenski:2003gt,Harmark:2008gm}. We believe this match for strings, as well as the match found for D-branes in this paper, are not coincidences. Indeed, in \cite{Harmark:2008gm} it is argued that one can take the $g_s N \rightarrow 0$ limit on the string theory side in a reliable way without entering the quantum string theory regime. We believe that one can make a similar argument also in the D-brane case presented here.

Independently on the discussion of the order-of-limits issue, one can build on the match between $\CN=4$ SYM and string theory found here for D-branes at first order in $g_s$ and examine the two theories at second and higher orders in $g_s$.

We discuss these and other perspectives further in Sections \ref{sec:stringsmt} and \ref{sec:discussion}.

\section{$SU(2)$ Spin Matrix Theory}

We consider $\CN=4$ SYM with gauge group $U(N)$ on $\R \times S^3$. 
Let $E$ be the energy, $S_1,S_2$ the angular momenta on $S^3$ and $J_1,J_2,J_3$ the three R-charges of the $SO(6) \simeq SU(4)$ R-symmetry, all measured in units of the inverse radius of $S^3$.

We consider the unitarity bound $E \geq J$ with $J=J_1+J_2$ \cite{Dobrev:1985qv}. The $SU(2)$ SMT limit is \cite{Harmark:2014mpa}
\begin{equation}
\label{SMT_limit}
H = J + \lim_{\lambda \rightarrow 0} \frac{g}{\lambda} (E-J)
\end{equation}
where $g$ is the coupling constant and $H$ the Hamiltonian of $SU(2)$ SMT. Any state with $E-J$ of order one or higher will decouple in this limit, hence the Hilbert space of SMT is greatly reduced. At the same time the interaction is simplified since all terms beyond one-loop go to zero.
We note that SMT at strong coupling $g \gg 1$ is dominated by states with small $(E-J)/\lambda$. 

Taking the limit \eqref{SMT_limit} one finds \cite{Harmark:2014mpa}
\begin{equation}
\label{H_SMT}
H = \tr ( a_1^\dagger a_1 + a_2^\dagger a_2 ) - \frac{g}{8\pi^2 N} \tr ( [a_1^\dagger, a_2^\dagger][a_1,a_2] )
\end{equation}
where $(a_s^\dagger)^i {}_j$ are the raising operators and $(a_s)^i {}_j$ the lowering operators with $s=1,2$ being the index of the spin $1/2$ representation of $SU(2)$ and $i,j=1,2,...,N$ the index for the adjoint representation of $U(N)$ (hence the name Spin Matrix Theory). We have $[(a_s)^j{}_i , (a_r^\dagger)^k {}_l ]= \delta_{s,r} \delta^k_i \delta^j_l$. The vacuum $|0\rangle$ of the Hilbert space is defined by $(a_s)^i {}_j |0\rangle=0$. The states in the Hilbert space are required to obey the singlet constraint
\begin{equation}
\label{singlet_cond}
\sum_{s=1}^2 \sum_{k=1}^N \Big[ (a_s^\dagger)^i{}_k(a_s)^k{}_j -  (a_s^\dagger)^k{}_j(a_s)^i{}_k \Big] |0\rangle =0
\end{equation} 
Thus, the Hilbert space is spanned by multitrace states. 

In the limit $N \rightarrow \infty$ with $g$ fixed the multitrace states become linearly independent and one can interpret the single trace parts as spin chains. The Hamiltonian \eqref{H_SMT} then reduces to the ferromagnetic $XXX_{1/2}$ Heisenberg spin chain. This is the analogue of the planar regime of $\CN=4$ SYM. 

In this paper we are interested in a different limit. Namely, we keep $N$ fixed and consider $J$ sufficiently large such that the Hamiltonian \eqref{H_SMT} becomes approximately classical. Using coherent states one finds the classical matrix model \cite{Harmark:2014mpa}
\begin{eqnarray}
\label{SMT_MM}
H  &=&  \frac{1}{2} \tr ( P_1^2 + P_2^2 + X_1^2 + X_2^2) \nn \\ &&
 - \frac{g}{32 \pi^2 N}\tr \Big(  [ X_1,X_2]^2  + [P_1,P_2]^2   + [X_1,P_1]^2 
 \nn \\ &&
 + [X_2,P_2]^2 + [X_1,P_2]^2 + [X_2,P_1]^2   \Big)
\end{eqnarray}
where $X_s$ and $P_s$ are $N$ by $N$ Hermitian matrices.
The singlet condition \eqref{singlet_cond} becomes the Gauss constraint
\begin{equation}
\label{gauss}
[X_1,P_1]+[X_2,P_2]=0
\end{equation}
For $g$ finite or small one needs $J \gg N^2$ where $J$ is the first line of \eqref{SMT_MM}. 
In the limit of infinite coupling $g=\infty$ the matrix model \eqref{SMT_MM} reduces to the diagonal components of the four matrices and one gets $H=J$. The Hamiltonian reduces to
\begin{equation}
\label{HequalJ}
H = \sum_{s=1}^2 \sum_{i=1}^N \Big[ ((P_s)^i {}_i)^2 +  ((X_s)^i {}_i)^2  \Big]
\end{equation}
which is $2N$ decoupled classical one-dimensional harmonic oscillators. This reproduces the counting of \cite{Kinney:2005ej} for $J \gg N^{3/2}$.  The states counted by \eqref{HequalJ} corresponds to $E=J$ states in $\CN=4$ SYM which are $1/4$ BPS.

\section{Giant gravitons}

We introduce now the relevant parameters on the string theory side and give a short summary of  giant gravitons. Let $R$ be the radius of $\ads_5$ and $S^5$, and $g_s$ and $l_s$ the string coupling and string length of type IIB string theory, respectively. We have $R^4 = 4\pi g_s N l_s^4$. Furthermore, $E$ is the energy, $S_1$ and $S_2$ the angular momenta on $\ads_5$ and $J_1,J_2,J_3$ the angular momenta on $S^5$, all measured in units of $1/R$. 

According to the dictionary of the AdS/CFT correspondence we have the relations $\gym^2 = 4\pi g_s$ and $R^4 = \lambda l_s^4$, while $E,S_1,S_2,J_1,J_2,J_3$ are identified between the two sides.

States with $E - J \ll 1$ in $\CN=4$ SYM are mapped to strings of type IIB string theory moving on the $S^5$ when $J$ does not grow with $N$. Instead when $J$ is proportional to $N$ the states are mapped to D-branes in the form of giant gravitons and when $J$ is proportional to $N^2$ they are mapped to modifications of the geometry.

A single giant graviton has $E = J=J_1+J_2$ of order $N$ and is $1/2$ BPS. It can either be an AdS giant graviton corresponding to a D3-brane wrapped on an $S^3$ inside $\ads_5$, or a sphere giant graviton corresponding to a D3-brane wrapped on an $S^3$ inside $S^5$, in both cases with angular momenta $J_1$ and $J_2$ on the $S^5$ \cite{McGreevy:2000cw}.
Several AdS or sphere giant gravitons with $E=J$ are generically $1/4$ BPS. The counting of such multiple giant graviton states is considered in \cite{Biswas:2006tj,Mandal:2006tk} matching that of \cite{Kinney:2005ej}. The spectrum corresponds to $N$ bosons in a two-dimensional harmonic oscillator potential which at high enough energies reduces to the spectrum of $2N$ decoupled harmonic oscillators, matching the spectrum of \eqref{HequalJ}.

Giant gravitons with $E=J$ are dual to restricted Schur polynomials built using Young tableaux with the maximal number of rows being $N$ since they correspond to representations of $U(N)$ \cite{Corley:2001zk,Balasubramanian:2004nb,deMelloKoch:2007rqf}. When mapping to sphere (AdS) giant gravitons one can interpret the columns (rows) as individual giant gravitons \cite{Corley:2001zk}. A Young tableau with many more rows than columns is dual to sphere giant gravitons, while a Young tableau with many more columns than rows is dual to AdS giant gravitons \cite{Bena:2004qv}. 
The maximal number of AdS giant gravitons is $N$ since each AdS giant graviton depletes the Ramond-Ramond five-form flux by one unit inside the three-sphere. This matches the fact that the maximal number of rows for a $U(N)$ representation is $N$.

\section{Proposal for match}

The goal of this paper is to match the classical matrix model \eqref{SMT_MM} to a corresponding matrix model found from D-branes of type IIB string theory on $\ads_5\times S^5$. The diagonal case Eq.~\eqref{HequalJ} can be treated using the DBI action for each D-brane. However, this does not capture the interaction term in Eq.~\eqref{SMT_MM}. For that we need to employ the non-abelian DBI action 
\cite{Myers:1999ps}. 

Below we consider $k$ AdS giant gravitons with $k\ll N$ corresponding to $k$ D3-branes wrapped on three-spheres inside $\ads_5$.  We cannot consider the maximal number $N$ of AdS giant gravitons since that would backreact on the background $\ads_5 \times S^5$ geometry and hence render the non-abelian DBI action invalid. However, for AdS giant gravitons, taking $k$ out of the $N$ D3-branes corresponds to breaking $U(N)$ into $U(k) \times U(N-k)$, where the $U(k)$ is the symmetry of the $k$ D3-branes and the $U(N-k)$ part corresponds to the D3-branes generating the $\ads_5\times S^5$ background with $N-k$ units of flux. 

Moreover, breaking $U(N)$ into $U(k) \times U(N-k)$ has a clear interpretation in $SU(2)$ SMT. Here it corresponds to exciting only the $U(k)$ part of the adjoint representation of $U(N)$. In the classical matrix model \eqref{SMT_MM} it means one turns on only a $k$ by $k$ matrix inside the full $N$ by $N$ matrix. The orthogonal $N-k$ by $N-k$ matrix not turned on then corresponds to the $\ads_5\times S^5$ background with $N-k$ units of flux. 

In conclusion we propose that the classical matrix model \eqref{SMT_MM} with Gauss constraint \eqref{gauss}
where $X_s$ and $P_s$ are $k$ by $k$ Hermitian matrices is dual to $k$ AdS giant gravitons at sufficiently high energy. 

For $k$ non-interacting AdS giant gravitons it is already clear that this proposal is correct since at high energies these have the spectrum of $2k$ harmonic oscillators, as found using the DBI action in \cite{Mandal:2006tk}.

\section{Interacting AdS giant gravitons from non-abelian DBI action}


For D3-branes the non-abelian DBI Lagrangian is \cite{Myers:1999ps}
\begin{eqnarray}
\label{NADBI}
\CL &=& T_{\rm D3}\, \mbox{STr} \Big( - \sqrt{ - \det ( g_{ab} + F_{aI} F_{bJ} M^{IJ} + 2\pi l_s^2 F_{ab} )} \nn \\ && \times \sqrt{ \det ( \delta^I_J + F^{IK} g_{KJ} )} + C_{0123} \Big) 
\end{eqnarray}
where we assumed the static gauge choice for the world-volume coordinates $x^a=\sigma^a$ and $x^I$ are the transverse coordinates that become $k$ by $k$ Hermitian matrices in the above Lagrangian. The field strength components are $F_{ab} = \partial_a A_b - \partial_b A_a + [A_a,A_b]$, $F_{aI} = g_{IJ} ( \partial_a x^I + i [A_a,x^I])$, $F^{IJ} = \frac{i}{2\pi l_s^2} [x^I , x^J]$ and we have defined $M^{IJ} = g^{IJ} + F^{IK}g_{KL} F^{LJ}$.
The tension is $T_{\rm D3} = 1/((2\pi)^3 g_s l_s^4)$. The above assumes a background where only the metric and the Ramond-Ramond four-form gauge potential $C_{(4)}$ is turned on. $\mbox{STr}(\cdots)$ means that one symmetrizes the expression in terms of the field strengths before taking the trace. The Lagrangian \eqref{NADBI} is valid up to $F^6$ terms. 

The metric for $\ads_5\times S^5$ is
\begin{equation}
\label{metric}
ds^2 = - ( r^2 + R^2 ) dt^2 + \frac{dr^2}{1 + \frac{r^2}{R^2}} + r^2 d\Omega_3^2 + dx^i dx^i
\end{equation}
where the five-sphere is defined by $x^2 = x^i x^i = R^2$ with $i=1,2,...,6$ and the three-sphere metric is $d\Omega_3^2 = d\psi^2 + \cos^2\psi \, d\chi^2 + \sin^2 \psi \, d\phi^2$. The embedding of the $k$ D3-branes is
$t= \sigma^0$, $\psi = \sigma^1$, $\chi=\sigma^2$, $\phi=\sigma^3$, $r=r(t)$ and $x^i = x^i(t)$ where the transverse coordinates are $x^I = (x^i , r)$. The transverse coordinates $x^i$ and $r$ are $k$ by $k$ Hermitian matrices. The five-sphere constraint is the $k$ by $k$ matrix equation $x^2 = R^2 I$. We set the world-volume gauge field to zero $A_a=0$ which gives us the Gauss constraint $[x^i , p_i ] + [r,p_r]=0$. Below we study the Lagrangian
\begin{equation}
\label{L2}
L = \int d\Omega_3 \CL -  \tr \left[    \frac{N \Lambda}{R^2}  \Big( x^2- R^2 \Big) \right]
\end{equation}
where we integrated \eqref{NADBI} over the unit three-sphere and added a Lagrange multiplier term to impose the $x^2 =R^2$ constraint.

Below we explore the Lagrangian \eqref{L2} in the following regime
\begin{itemize}
\item We take the limit of large radius $r \gg R$. This will serve both as a limit in which AdS giant gravitons dominate over sphere giant gravitons and as a high energy limit.
\item We consider only states with $E- J \ll 1$ since these are the states dual to the $SU(2)$ Spin Matrix Theory limit. 
\item We consider only the leading part of the interaction term between the D3-branes in the weakly interacting limit. Thus, we keep only the leading part involving the transverse field strength $F^{ij}$.
\end{itemize}

\subsection{Decoupling of radial modes}

Consider the limit of zero interactions between the k AdS giant gravitons. Then $r$ and $x^i$ are diagonal matrices. For large $r/R$ we find
\begin{eqnarray}
\label{diagL}
L &=& \tr \left[ - \frac{Nr^3}{R^3} \sqrt{ 1+ \frac{r^2}{R^2} - \frac{\dot{r}^2}{r^2+R^2} - \frac{\dot{x}^2}{R^2}} +\frac{N r^4}{R^4}  \right] \nn \\ & \simeq & \tr \left[ \frac{N \dot{r}^2}{2R^2}  + \frac{Nr^2 \dot{x}^2}{2R^4} - \frac{Nr^2}{2R^2}\right]
\end{eqnarray}
where dot means the time derivative. The corresponding Hamiltonian is
\begin{equation}
\label{diagH}
E = \tr \left[ \frac{R^2 p_r^2}{2N} + \frac{R^4 p^2}{2N r^2} + \frac{Nr^2}{2R^2} \right]
\end{equation}
We can ignore the Lagrange multiplier term in \eqref{L2} since we focus below on the dynamics of $r$.
The Hamiltonian \eqref{diagH} has a minimum at $N r^4 = R^6 p^2$. Writing $N r_{\rm min}^4 = R^6 p^2$ and $z=\sqrt{N}(r-r_{\rm min})/R$ we find near the minimum $H = \tr ( p_z^2 + 2z^2 + Nr_{\rm min}^2 /R^2 )$. This corresponds to $k$ harmonic oscillators of mass $2$. Hence if any of these were excited $E-J$ would be of order one or higher (below we shall see that $\tr[Nr^2/(2R^2)]$ is cancelled by subtracting $J$). Therefore, the radial modes for the D3-branes are decoupled, also in the weakly interacting case. Thus, we set $\dot{r}=0$ from now on. This means we should impose $\partial L / \partial r = 0$ as a constraint. Considering \eqref{diagL} we see this gives $\dot{x}^2 =R^2$ hence in the weakly interacting case $\dot{x}^2 \sim R^2$. 

Define $r_0= \tr(r) /k$. As stated above, we are considering the regime $r_0 \gg R$.
In the weakly interacting case we take the eigenvalues of $r$ to be of order $r_0$, $i.e.$ that the deviation of each eigenvalue from $r_0$ is much smaller than $r_0$. This means that
\begin{equation}
\label{Fri}
\frac{1}{r_0^2 R^2} | \tr(F^{ri} F^{ri}) | \ll \frac{1}{R^4} | \tr ( F^{ij}F^{ij} ) |
\end{equation}

\subsection{Expansion of Lagrangian}

From the metric \eqref{metric} we see that the proper time approximately is $\tau = r_0 t$. Hence the velocities of the $k$ AdS giant gravitons are approximately 
\begin{equation}
\Big( \frac{dx}{d\tau} \Big)^2 \sim \frac{\dot{x}^2}{r_0^2} \sim \frac{R^2}{r_0^2} \ll 1
\end{equation}
Hence we are in a regime with small velocities. Thus, we should expand the Lagrangian \eqref{L2} in powers of $F^{aI}$. Moreover, we should also expand in powers of $F^{IJ}$ as well since we are in the weakly interacting limit. The leading order Lagrangian is thus obtained by considering the terms quadratic in the field strengths. We find
\begin{equation}
\label{L3}
L = \tr \left[  \frac{Nr^2}{2R^4} ( \dot{x}^2 - R^2)  -   \frac{N \Lambda}{R^2}  \Big( x^2- R^2 \Big) - \frac{N r^4}{4R^4} F^2 \right]
\end{equation}
In deriving this result one finds that the zeroth order term in the field strength expansion cancels out with the $C_{0123}$ term. Notice that we omitted the symmetrized trace prescription. This is due to the fact that we can effectively regard $r$ and $x^i$ as commuting variables to the order we are working. First we notice that $F^2 = F^{ij} F^{ij}$ to leading order in the interactions due to \eqref{Fri} since $F^{IJ} F_{IJ} = F^{ij} F^{ij} + \frac{2}{1+r^2/R^2} F^{ri} F^{ri}$. Secondly, the difference between $\tr ((r\dot{x})^2)$ and $\tr (r^2 \dot{x}^2)$ goes like $\tr ( [r,\dot{x}^i]^2 )$ which is much smaller than $\tr ( r^4 F^2 )$.

From $\partial L / \partial r=0$ we get the constraint 
\begin{equation}
\label{rconst}
\dot{x}^2 = R^2 + r^2 F^2
\end{equation}
Computing the equations of motion (EOMs) for the five-sphere directions we find 
\begin{equation}
\label{transEOM}
\ddot{x}^i = - \frac{2R^2}{r^2} \Lambda x^i - \frac{r^2}{4} \frac{\partial F^2}{\partial x^i} 
\end{equation}
with $(2\pi l_s^2)^2 \frac{\partial F^2}{\partial x^i} = - 4  [x^j,[x^i,x^j]]$. Using now $x^2=R^2$ we find $\dot{x}^2 = - x^i \ddot{x}^i = - \ddot{x}^i x^i$ ($[x^i,\ddot{x}^i]=0$ due to the Gauss constraint) and $x^i  \frac{\partial F^2}{\partial x^i} = \frac{\partial F^2}{\partial x^i} x^i  = 4 F^2$. Thus, contracting \eqref{transEOM} with $-x^i$ we get \eqref{rconst} provided we set
\begin{equation}
\Lambda = \frac{r^2}{2R^2}
\end{equation}
to be the on-shell value of $\Lambda$. Hence with this the EOMs \eqref{transEOM} imply the constraint \eqref{rconst}.

We can now write the Lagrangian \eqref{L3} as
\begin{equation}
\label{L4}
L(x^i , \dot{x}^i) = \tr \left[  \frac{Nr^2}{2R^4}  \dot{x}^2   -   \frac{N r^2}{2R^4}  x^2 - \frac{N r^4}{4R^4} F^2  \right]
\end{equation} 
with $x^2 = R^2$ to be imposed on the EOMs. Making the coordinate transformation
$y^i = \sqrt{N} r x^i / R^2$ this can be written as 
\begin{equation}
\label{L5}
L(y^i , \dot{y}^i) = \tr \left[  \frac{1}{2}  \dot{y}^2   -   \frac{1}{2}  y^2 - \frac{R^4}{4N} \hat{F}^2  \right]
\end{equation} 
where $\hat{F}^{jk} = i [y^j , y^k] /(2\pi l_s^2)$. The constraint is $y^2 = N r^2/R^2$. Since $r$ does not appear in the Lagrangian this is just expressing that $y^2$ is constant. Hence we can write this constraint as $y^i \dot{y}^i=0$. The corresponding Hamiltonian description is
\begin{equation}
\label{Ey}
E(y^i,\hat{p}_i) = \tr \left[  \frac{1}{2}  \hat{p}^2   +   \frac{1}{2}  y^2 + \frac{R^4}{4N} \hat{F}^2  \right]
\end{equation} 
along with the constraints $y^i \hat{p}_i=0$ and $[y^i,\hat{p}_i] = 0$ where we defined the momenta $\hat{p}_i = \dot{y}^i$. 

\subsection{Spin Matrix Theory regime}

We have $J_1 = \tr ( y^1 \hat{p}_2 - y^2 \hat{p}_1 )$ and $J_2 = \tr ( y^3 \hat{p}_4 - y^4 \hat{p}_3 )$.
Make now the canonical transformation of the Hamiltonian $E(y^i,\hat{p}_i)$
\begin{equation}
\begin{array}{c}
y_1 = \frac{1}{\sqrt{2}} ( P_1 + X_3  ) \spa y_2 = \frac{1}{\sqrt{2}} ( X_1 + P_3  ) 
\\[3mm]
 y_3 = \frac{1}{\sqrt{2}} ( P_2 + X_4  ) \spa y_4 = \frac{1}{\sqrt{2}} ( X_2 + P_4  )
\\[3mm]
\hat{p}_1 = \frac{1}{\sqrt{2}} ( -X_1 + P_3  ) \spa \hat{p}_2 = \frac{1}{\sqrt{2}} ( P_1 - X_3  ) 
\\[3mm]
 \hat{p}_3 = \frac{1}{\sqrt{2}} ( -X_2 + P_4  ) \spa \hat{p}_4 = \frac{1}{\sqrt{2}} ( P_2 - X_4  )
\end{array}
\end{equation}
We find $E-J = \tr ( P_3^2 + P_4^2 + X_3^2 + X_4^2 + ( \hat{p}_5^2 + \hat{p}_6^2 + (y^5)^2+ (y^6)^2 ) + R^4 F^2 /(4N) )$. We see that any excitation of $X_3$, $X_4$, $y^5$ or $y^6$ would give $E-J$ of order one or higher. Hence these modes decouple when considering the Spin Matrix regime $E - J \ll 1$. Thus, $X_3=X_4=y^5=y^6=0$ and $P_3=P_4=\hat{p}_5=\hat{p}_6=0$. Inserting this in \eqref{Ey} we find for the surviving modes that the Hamiltonian $E(X_s,P_s)$, $s=1,2$, is given by
\begin{eqnarray}
\label{string_MM2}
E  &=&  \frac{1}{2} \tr ( P_1^2 + P_2^2 + X_1^2 + X_2^2) \nn \\ &&
 - \frac{g_s}{8\pi}\tr \Big(  [ X_1,X_2]^2  + [P_1,P_2]^2   + [X_1,P_1]^2 
 \nn \\ &&
 + [X_2,P_2]^2 + [X_1,P_2]^2 + [X_2,P_1]^2   \Big)
\end{eqnarray}
subject to the Gauss constraint $[X_1,P_1]+[X_2,P_2]=0$, the same as anticipated in Eq.~\eqref{string_MM}.
We note that the constraint $y^i \hat{p}_i = 0$ is identically satisfied. We have $J = \frac{1}{2} \tr ( P_1^2 + P_2^2 + X_1^2 + X_2^2)$.

The Hamiltonian \eqref{string_MM2} is valid for $r \gg R$ which is equivalent to $J \gg k N$. Clearly this is a high-energy regime in which the AdS giant gravitons dominate over the sphere giant gravitons, since the dual states of $SU(2)$ SMT correspond to Young tableaux that typically have $k$ rows and more than $N$ columns. Moreover, the result \eqref{string_MM2} is valid for $g_s \ll 1$ up to $\CO(g_s^2)$.

Comparing \eqref{string_MM2} to the $SU(2)$ SMT result \eqref{SMT_MM} with $X_s$ and $P_s$ being Hermitian $k$ by $k$ matrices, we see that it is the exact same matrix model. In detail, one translates \eqref{SMT_MM} to $\CN=4$ SYM by replacing $g$ with $\lambda=\gym^2 N$ and $H$ with $E$, as seen from \eqref{SMT_limit}. The match between $\CN=4$ SYM and type IIB string theory is then found using $\gym^2 = 4\pi g_s$.

The above result is stable under the following extensions/modifications:
\begin{itemize}
\item We put $k$ D3-branes in the background with $N$ units of flux. Inside the $k$ spheres the flux is therefore $N-k$ units. Is the result still valid if one puts another $k' \ll N$ D3-branes in the background with $N-k$ units of flux? This is indeed the case, since the matrix model \eqref{string_MM2} does not depend on $N$. Indeed, one could even combine the $k+k'$ D3-branes in the same non-abelian DBI action by making the AdS radius $R$ into a $k+k'$ by $k+k'$ matrix and replacing $N$ with $R^4 / ( 4\pi g_s l_s^4 )$. 
\item For simplicity we compared the matrix model \eqref{string_MM2} to the high energy limit of $SU(2)$ SMT in the adjoint representation of $U(N)$. The above match still holds if one instead considers the adjoint representation of $SU(N)$ where $X_s$ and $P_s$ are traceless $N$ by $N$ matrices. In this case $SU(N)$ breaks into $U(k) \times SU(N-k)$. Thus, one gets the same matrix model from $SU(2)$ SMT in the adjoint representation of $U(k)$ to match \eqref{string_MM2}. 
\end{itemize}

In conclusion, we have found a match between the classical limit of strongly coupled $SU(2)$ SMT, corresponding to $\CN=4$ SYM close to the unitarity bound $E\geq J$, and the dynamics of interacting AdS giant gravitons on $\ads_5\times S^5$.

\section{Connecting string theory to SMT}
\label{sec:stringsmt}

As summarized in the Introduction, the matrix model Eq.~\eqref{string_MM2} is found from $SU(2)$ SMT with $g_s \ll 1/N$ while on the string side $1/N \ll g_s \ll 1$. We believe it is not a coincidence that one gets the same matrix model, and that one should be able to connect the two regimes. 

We first remark that the one-loop correction in weakly coupled $\CN=4$ SYM is believed to be special for supersymmetric states. Indeed, it is conjectured \cite{Kinney:2005ej} that if one uses the tree-level plus one-loop dilatation operator to find supersymmetric states of $\CN=4$ SYM, no further reduction of these states will occur at higher loops. By the AdS/CFT correspondence, this is equivalent to asserting that SMT for $g=\infty$ is dual to the supersymmetric states of type IIB string theory with $E=J$ \cite{Harmark:2014mpa}. 

Can one extend this to a duality between SMT for $g\gg 1$ and type IIB string theory in the SMT limit beyond the supersymmetric states? With the results of this paper, we have now two very different regimes of $SU(2)$ SMT where this is the case. In addition to the regime where the matrix model \eqref{string_MM2} is valid, we have the planar regime as well. This corresponds to taking the $N\rightarrow \infty$ limit of $SU(2)$ SMT. $SU(2)$ SMT then becomes the ferromagnetic $XXX_{1/2}$ Heisenberg spin chain acting on single-traces in the Hilbert space. For $g \gg 1$ the low energy dynamics of the spin chain dominates. For $J\gg 1$ one finds a spectrum of magnons for which $H-J \sim g/J^2$. The classical limit with high quantum numbers of the magnons, corresponding to $H-J \sim g /J$, is described by the Landau-Lifshitz sigma-model. In $\CN=4$ SYM this translates to the regime $E-J \sim \lambda / J$ with $\lambda \ll 1$ and $J \gg 1$. 

On the string side, the same Landau-Lifshitz model is found in the regime $E -J \ll 1$ to first order when expanding $\lambda / J^2 \ll 1$ with $\lambda \gg 1$ \cite{Kruczenski:2003gt}, explaining the matchings of particular semi-classical operators to string states \cite{Frolov:2003qc}.

In \cite{Harmark:2008gm} it is argued that this match is not a coincidence. Indeed, starting from the string sigma-model on $\ads_5\times S^5$, it is argued that one can take the limit
\begin{equation}
\label{SMT_limit2}
H = J + \lim_{g_s \rightarrow 0} \frac{g}{4\pi g_s N} (E-J)
\end{equation}
and obtain the above-mentioned classical Landau-Lifshitz model, assuming that $J \gg 1$. The reasons this works are: 1) The $\ads_5\times S^5$ background is exact \cite{Kallosh:1998qs}. 2) For large $J$ the sigma-model action remains large and one is thus not in a quantum string regime. 3) The modes that decouple in the limit become infinitely heavy. 4) Zero-mode fluctuations are suppressed since $E=J$ is $1/4$ BPS. 

For the match found by this paper, we believe that one can argue along the same lines that the limit \eqref{SMT_limit2} of the non-abelian DBI action is possible. It is clear that by considering $J \gg kN$ the action is always large, and one can easily check that the higher-order field strength corrections in the action are suppressed when taking the limit $g_s \rightarrow 0$. Moreover, we have seen in this paper that the modes that decouple become infinitely heavy in the limit \eqref{SMT_limit2}. It would be interesting to examine these arguments  more closely. 

In conclusion, we have found two classical regimes with $J$ large where one can match strongly coupled $SU(2)$ SMT to type IIB string theory on $\ads_5\times S^5$, one with $J \sim N^0$ and one with $J\sim N$.  It would be highly interesting to see if it is also possible to find regimes of SMT for which a match with $J\sim N^2$ is possible.

\section{Discussion and conclusion}
\label{sec:discussion}

We have shown in this paper that we can match the strongly coupled limit of $SU(2)$ SMT at high energies to the dynamics of $k$ interacting AdS giant gravitons in the regime $J \gg kN$.  
This means we are able to match non-supersymmetric dynamics of D-branes on $\ads_5\times S^5$ to a finite-$N$ regime in $\CN=4$ SYM near a unitarity bound.
Thus, we are matching $\CN=4$ SYM in a regime where $J$ goes like $N$ to the dynamics of non-perturbative objects in string theory described by the non-abelian DBI action.
This is in contrast to previous results of 
\cite{Balasubramanian:2002sa} 
where one matches the dispersion relation of open strings ending on D-branes in the large $N$ limit to $\CN=4$ SYM. In that case one focusses on matching the dynamics of open strings with non-trivial boundary conditions provided by the giant gravitons. Hence, the giant gravitons are not themselves dynamical.

It could be interesting to extend our computation to second order in $\gym^2 = 4\pi g_s$. On the $SU(2)$ SMT side, this would correspond to a perturbation from the two-loop dilatation operator of $\CN=4$ SYM. On the string theory side, one should consider higher order corrections from the non-abelian DBI action, including $F^4$ terms for the transverse field strength. This could possibly provide a starting point to study the structure of higher-order corrections away from the SMT regime considered in this paper.

Regarding our use of the non-abelian DBI action \cite{Myers:1999ps} 
 in this paper, we note that to our knowledge it has only been employed in very few cases to the AdS/CFT correspondence \cite{Erdmenger:2007vj}. The results of this paper could possibly give a new way to explore and test non-abelian DBI also at order $F^6$ and beyond. 
Notice furthermore that the regime $r\gg R$ corresponds effectively to the matrix theory limit of the non-abelian DBI action \cite{Taylor:1997dy} since we found that velocities are small in this regime. Moreover, for $r \gg R$ the curvature of the embedding geometry of the D3-branes is small.

It is interesting to check the much simpler SMT limit coming from considering the unitarity bound $E \geq J_1$ using the results of this paper. Taking the limit \eqref{SMT_limit} in this case one finds $H = \tr ( a^\dagger a )$ with a singlet constraint on the spectrum \cite{Berenstein:2004kk}. Notice that there is no interaction term in this case which is connected to the fact that it describes the $1/2$ BPS sector of $\CN=4$ SYM. In the classical limit one gets the matrix model $H = \frac{1}{2} \tr ( P^2 + X^2 )$ with Gauss constraint $[X,P]=0$. Considering the string theory side, using the methods of this paper, it is not hard to see that one finds the Hamiltonian \eqref{string_MM} with $X_2=P_2=0$. Imposing then the Gauss constraint the interaction term vanishes, and hence one finds perfect agreement.

The $SU(2)$ SMT is the simplest non-trivial SMT one can consider. Considering instead the unitarity bounds $E \geq J_1+J_2+J_3$, $E \geq S_1 + J_1$, $E \geq S_1+S_2+J_1$ and $E\geq S_1+ S_2 + J_1 + J_2 + J_3$, respectively, one finds the $SU(2|3)$ SMT, $SU(1,1|2)$ SMT, $SU(1,2|2)$ SMT and $SU(1,2|3)$ SMT \cite{Harmark:2006di,Harmark:2014mpa}. 
It could be highly interesting to extend our match for interacting giant gravitons to these SMTs as well. 
$SU(1,2|3)$ SMT is particularly interesting since it contains a semi-classical configuration dual to a supersymmetric black hole with $E = S_1+ S_2 + J_1 + J_2 + J_3$ \cite{Gutowski:2004yv} that so far has eluded understanding 
\cite{Kinney:2005ej,Grant:2008sk}.

\section*{Acknowledgments}

We thank Jelle Hartong, Cindy Keeler, Niels Obers and Marta Orselli for nice discussions and useful comments on the draft. We acknowledge support from the Marie-Curie-CIG grant ``Quantum Mechanical Nature of Black Holes" from the European Union. We thank the Physics Department of Perugia University for kind hospitality while this research was carried out.



%

\end{document}